\documentclass[runningheads]{llncs}
\usepackage{graphicx}
\usepackage{cite}
\usepackage[font=small,skip=5pt]{caption}
\usepackage{booktabs}

\usepackage{amsmath,amssymb,amsfonts}

\usepackage{orcidlink} 

\begin{document}
\title{Automating Cloud Security and Forensics \\ Through a Secure-by-Design GenAI Framework}
\author{Dalal Alharthi\inst{1}\orcidID{0000-0003-0299-024X} \\
Ivan Roberto Kawaminami Garcia\inst{1}\orcidID{0000-0002-5206-4693}}

\authorrunning{D. Alharthi and I. Garcia}

\institute{University of Arizona, Tucson, AZ, USA\\
\email{\{dalharthi,kawaminami\}@arizona.edu}}

\maketitle  

\let\thefootnote\relax
\footnotetext{This is a preprint of a paper accepted to the International Conference on Digital Forensics and Cyber Crime (ICDF2C 2025).}

\begin{abstract}
As cloud environments become increasingly complex, cybersecurity and forensic investigations must evolve to meet emerging threats. Large Language Models (LLMs) have shown promise in automating log analysis and reasoning tasks, yet they remain vulnerable to prompt injection attacks and lack forensic rigor. To address these dual challenges, we propose a unified, secure-by-design GenAI framework that integrates PromptShield and the Cloud Investigation Automation Framework (CIAF). PromptShield proactively defends LLMs against adversarial prompts using ontology-driven validation that standardizes user inputs and mitigates manipulation. CIAF streamlines cloud forensic investigations through structured, ontology-based reasoning across all six phases of the forensic process. We evaluate our system on real-world datasets from AWS and Microsoft Azure, demonstrating substantial improvements in both LLM security and forensic accuracy. Experimental results show PromptShield boosts classification performance under attack conditions, achieving precision, recall, and F1 scores above 93\%, while CIAF enhances ransomware detection accuracy in cloud logs using Likert-transformed performance features. Our integrated framework advances the automation, interpretability, and trustworthiness of cloud forensics and LLM-based systems, offering a scalable foundation for real-time, AI-driven incident response across diverse cloud infrastructures.


\end{abstract}
\section{Introduction} \label{intro}

Large Language Models (LLMs) have demonstrated remarkable advancements across diverse applications, including cloud security and digital forensics. Their ability to mimic human reasoning enables automation in threat detection and incident response \cite{derner2024taxonomy,chernyshev2024forensic}. However, two persistent challenges hinder their safe and scalable deployment in mission-critical environments: the continued reliance on manual, error-prone forensic analysis, and the vulnerability of LLM-powered systems to adversarial prompt injection attacks that can manipulate model outputs and compromise security \cite{owasp2025promptinjection,akula2024cloudtools,purnaye2022cloudforensics}. While research has primarily emphasized LLM scalability and performance, their potential role in enhancing cloud forensic investigations remains underexplored. Cloud environments, in particular, continue to be susceptible to ransomware attacks that exploit misconfigurations and weak security policies. These attacks not only disrupt operations but also complicate forensic investigations through techniques such as encryption and obfuscation \cite{akula2024cloudtools,reshmi2021information}. Recent studies have systematically categorized the challenges of cloud forensics and analyzed evolving adversarial attack patterns \cite{purnaye2022cloudforensics,mishra2012cloudforensics}. Although some automated forensic analysis tools have emerged \cite{zhu2023promptbench,schulhoff2023hackaprompt}, they remain largely reactive and lack mechanisms for structured, ontology-driven validation. Meanwhile, AI-driven frameworks such as LangGraph \cite{langgraph}, AutoGen \cite{autogen}, and CrewAI \cite{crewAI} represent promising advancements in multi-agent coordination, yet their applications in forensic contexts are still limited.

To address these gaps, we introduce a unified, secure-by-design framework that integrates automation and proactive security into LLM-based cloud forensic workflows. This work presents two complementary components: (1) the \textbf{Cloud Investigation Automation Framework (CIAF)}, designed to automate cloud forensic log analysis using semantic validation and structured templates; and (2) \textbf{PromptShield}, a security-by-design framework that mitigates prompt injection attacks through ontology-driven prompt validation. While CIAF enhances forensic investigation efficiency and accuracy by standardizing input structure, PromptShield increases system robustness by replacing ambiguous or adversarial prompts with deterministic, expert-validated alternatives. Recent studies have emphasized the importance of structured methodologies to improve forensic accuracy and automate investigative processes \cite{al-mugerrn2023metamodeling,mishra2012cloudforensics}. By framing forensic log analysis within the context of causal reasoning and structured AI validation, CIAF establishes a scalable and interpretable foundation for cloud forensic investigations. The effectiveness of ontology-driven forensic analysis lies in its ability to impose structured constraints on cloud logs, reducing the noise and ambiguity often present in unstructured datasets \cite{zawoad2013cloudforensics}. This approach is consistent with adversarial robustness frameworks \cite{madry2018towards,carlini2017evaluating}, where structured constraints help reduce attack vectors and enhance forensic accuracy by filtering out irrelevant or misleading log events. Moreover, enforcing causal dependencies between attack patterns and forensic outcomes enables systematic validation through causal inference frameworks \cite{pearl2009causality}. Understanding these interactions is critical for quantifying investigative effectiveness and assessing generalization trade-offs in AI-driven environments \cite{zhou2024algorithmic,li2023latent}.

Building on these insights, this paper introduces a dual-layered solution for cloud forensics and LLM security. Our main contributions are as follows:
(1) We design and implement a dual-layered architecture that bridges LLM security and cloud forensic automation through semantic validation. (2) We develop PromptShield to standardize and secure LLM inputs, achieving high resilience to adversarial manipulation with precision, recall, and F1 scores exceeding 93\%. (3) We evaluate CIAF through a real-world ransomware case study on Microsoft Azure, demonstrating significant improvements in forensic accuracy and interpretability using Likert-scaled performance features. (4) We provide experimental validation across both AWS and Azure environments, showing that our framework enhances the effectiveness, reliability, and scalability of AI-driven forensic investigations.

The remainder of this paper is structured as follows: Section~\ref{related-work} discusses related work on LLM security, ontology-driven forensics, and ransomware detection. Section~\ref{framework} presents the proposed framework in detail. Section~\ref{experiments} describes the experimental setup and evaluation results. Section~\ref{discussion} outlines key insights and future research directions. Finally, Section~\ref{conc} concludes the paper.


\section{Related Work and State of the Art} \label{related-work}

\textbf{\textit{Foundations of Adversarial Robustness and Privacy in AI.}}
Before the rise of AI-driven cloud forensics, research primarily focused on traditional forensic methodologies, which relied on manual log analysis and rule-based approaches to investigate cyber incidents. Foundational work by Goodfellow et al. \cite{goodfellow2015explaining} introduced adversarial examples, demonstrating how small perturbations in input data could cause deep learning models to misclassify. Building on this, Carlini and Wagner \cite{carlini2017towards} developed stronger attack methods and evaluated countermeasures, revealing persistent vulnerabilities in deep networks. In parallel, advances in adversarial robustness focused on certified defenses, such as randomized smoothing \cite{cohen2019certified}, which provides probabilistic guarantees of model resilience under adversarial perturbations. Privacy concerns also emerged as a critical research area, with Differential Privacy \cite{dwork2006calibrating} establishing formalized mechanisms to protect data while maintaining utility. These foundational studies set the stage for evolving research into the vulnerabilities of complex, high-dimensional ML systems. As scaling continues to drive AI performance, recent work suggests that structured learning approaches offer alternative pathways to enhancing security \cite{snell2024scaling}. These foundational efforts paved the way for the emergence of structured, AI-driven methodologies in cybersecurity and digital forensics, which we discuss in the following subsections.

\textit{\textbf{LLMs in Cybersecurity and Digital Forensics.}}
With the rise of LLMs and Generative AI, new security risks have emerged, particularly prompt injection attacks, which manipulate the natural language flexibility of LLMs to produce unintended outputs. Recent work has systematically evaluated these attacks, highlighting their systemic risks in multi-agent settings \cite{liu2023prompt,liu2024automatic}. In multi-agent LLM environments, research has shown that manipulated prompts can propagate cascading failures, affecting autonomous decision-making in critical infrastructures such as transportation networks and cloud security systems \cite{ju2024flooding}. Existing mitigation strategies highlight the importance of structured defenses in distributed environments \cite{zhang2022security}. However, ensuring robust security in large-scale, collaborative AI deployments remains a significant challenge, requiring a deeper integration of theoretical guarantees for adversarial robustness and causality-aware security frameworks \cite{muliarevych2024security}. Recent studies have demonstrated the growing role of LLMs in cybersecurity, particularly in automating security analysis and enhancing forensic readiness. Yao et al. \cite{yao2025llms} provided a systematic review of LLM applications in cybersecurity, highlighting their ability to process complex security logs, detect anomalies, and generate actionable forensic insights. As LLMs continue to be integrated into forensic workflows, ensuring the integrity and auditability of their outputs becomes a crucial area of research \cite{bendiab2024forensic,galhotra2024logging}.

\textbf{\textit{Ontology-Driven Cloud Forensics.}}
Traditional cloud forensics has long faced challenges in processing large volumes of unstructured log data. Security Information and Event Management (SIEM) systems have been widely used for detecting suspicious activities by correlating log data from various sources \cite{denning1987intrusion,kent2006guide}. However, these rule-based methods often struggled to keep pace with evolving cyber threats and required significant manual intervention to update rules and models. Ontology-based frameworks have been proposed to address this challenge by providing structured representations of forensic data, improving accuracy and efficiency in forensic investigations. For example, Rouached et al. \cite{rouached2018ontology} introduced an ontology-driven approach for web services logs, enabling better cyber-attack detection. Similarly, the Cloud Forensic Readiness as a Service (CFRaaS) model \cite{li2023cloudforensic} emphasizes proactive accumulation of digital evidence, leveraging ontological structures to enhance forensic readiness in cloud environments. Beyond forensic readiness, ontology engineering has become a critical tool in cyber threat intelligence. A recent study by Bratsas et al. \cite{bratsas2024knowledge} highlights how knowledge graphs and semantic web tools can improve forensic investigations by structuring cyber threat data. This approach reduces investigative ambiguity, enhances forensic accuracy, and ensures a standardized methodology for detecting and responding to cyber incidents.

\textbf{\textit{AI-Based Ransomware Detection and Cloud Security.}}
Despite advancements in forensic automation, cloud environments remain highly vulnerable to ransomware attacks, which exploit misconfigurations and weak security policies \cite{akula2024cloudtools}. Ransomware not only disrupts operations but also complicates forensic investigations by leveraging encryption and obfuscation techniques that hinder log analysis and evidence collection \cite{reshmi2021information}. Recent studies have systematically categorized cloud forensics challenges and analyzed adversarial attack patterns \cite{purnaye2022cloudforensics,mishra2012cloudforensics}. While automated forensic analysis tools exist \cite{zhu2023promptbench,schulhoff2023hackaprompt}, they remain reactive and lack structured ontology-driven validation. Emerging AI-driven frameworks, such as LangGraph \cite{langgraph}, AutoGen \cite{autogen}, and CrewAI \cite{crewAI}, offer promising advancements, but their forensic applications remain limited. The application of AI-based ransomware detection is rapidly evolving. Ahmad et al. \cite{ahmad2025ai} conducted a comprehensive review of AI-driven techniques for identifying ransomware in cloud environments, emphasizing the need for anomaly detection and predictive modeling. In addition, machine learning (ML) approaches have been explored for forensic automation. Alhawi et al. \cite{alhawi2025trusted} proposed a meta-feature-based detection method leveraging volatile memory analysis to improve ransomware identification in private cloud infrastructures. These studies demonstrate how AI-powered solutions can proactively detect and mitigate ransomware attacks before they cause significant damage.

\textbf{\textit{LLM-Based Cloud Forensic Investigation Tools.}}
The integration of LLMs into digital forensics is an emerging field, with studies showing their potential to automate forensic processes and improve investigative accuracy \cite{bendiab2024forensic}. A recent study highlights how LLM invocation logging enhances forensic readiness, ensuring transparency and auditability in cloud investigations \cite{galhotra2024logging}. Despite these advancements, challenges persist in ensuring the integrity and reliability of cloud logs, which are critical for forensic investigations \cite{mishra2012cloudforensics}. Additionally, as cloud environments grow more complex, forensic methodologies must evolve to address emerging threats and vulnerabilities \cite{zawoad2013cloudforensics}. New research has explored LLM-based automation for cloud forensic investigations. LLMCloudHunter \cite{schwartz2024llmcloudhunter} introduces a framework that utilizes Large Language Models (LLMs) to automatically generate detection rules from unstructured cloud-based cyber threat intelligence (CTI) sources. Similarly, LogPrécis \cite{boffa2023logprecis} applies LLMs to Unix shell attack logs, extracting attacker tactics and reducing large datasets into manageable forensic fingerprints for enhanced investigation. These studies demonstrate the growing role of LLMs in automating forensic analysis, significantly improving the speed and efficiency of threat detection.

\textbf{\textit{Beyond Forensics: AI in Cybersecurity Operations.}}
Beyond forensic analysis, AI-enhanced Risk-Based Access Management (RBAM) systems have been developed to dynamically adjust permissions by analyzing access logs, leading to reduced false positives and unauthorized access incidents \cite{agorbia2024leveraging}. Additionally, SecGenAI, a framework for securing cloud-based GenAI applications, has been proposed to enhance privacy compliance and mitigate adversarial threats in cloud security \cite{secgenai_cloud_genai_security}. However, while these advancements underscore the potential of GenAI in cybersecurity, they also raise concerns regarding its dual-use nature—where the same technology can be exploited for malicious purposes \cite{dual_use_genai_cybersecurity}. As AI and LLM-based forensic solutions evolve, threat modeling remains a key strategy in mitigating cybersecurity risks. Threat modeling is a structured approach to identifying, assessing, and mitigating security threats to a system, application, or network. It involves defining assets, recognizing potential threats, analyzing attack vectors, assessing risks, and implementing security controls \cite{verma2024operationalizingthreatmodelredteaming}. This process is particularly crucial in LLM-driven cloud forensics, as AI systems introduce unique vulnerabilities that must be accounted for in forensic investigations.


\section{Proposed Framework} \label{framework}
Our proposed framework integrates two critical dimensions of cybersecurity: automated cloud forensic investigations and proactive large language model (LLM) security. It combines the structured forensic process with a security-by-design philosophy tailored to the unique challenges of modern cloud and AI systems. At its core, the framework adheres to the six-step cloud forensic process: identification of an event, identification of evidence, collection of evidence, analysis of evidence, interpretation of results, and presentation of results. The process begins with detecting suspicious activities or anomalies using monitoring systems. Advanced forensic tools are applied to analyze the data, uncover patterns, reconstruct attack timelines, and interpret the findings to determine causes, responsible entities, and the scope of the impact. Finally, findings are compiled into a clear, comprehensive report that is suitable for stakeholders and potential legal proceedings. Although traditional forensic processes are executed manually. Our framework automates the entire pipeline to improve speed, accuracy, and scalability in response to cyberattacks in cloud environments.

\begin{figure}[t]
    \begin{center}
    \centerline{\includegraphics[width=\columnwidth]{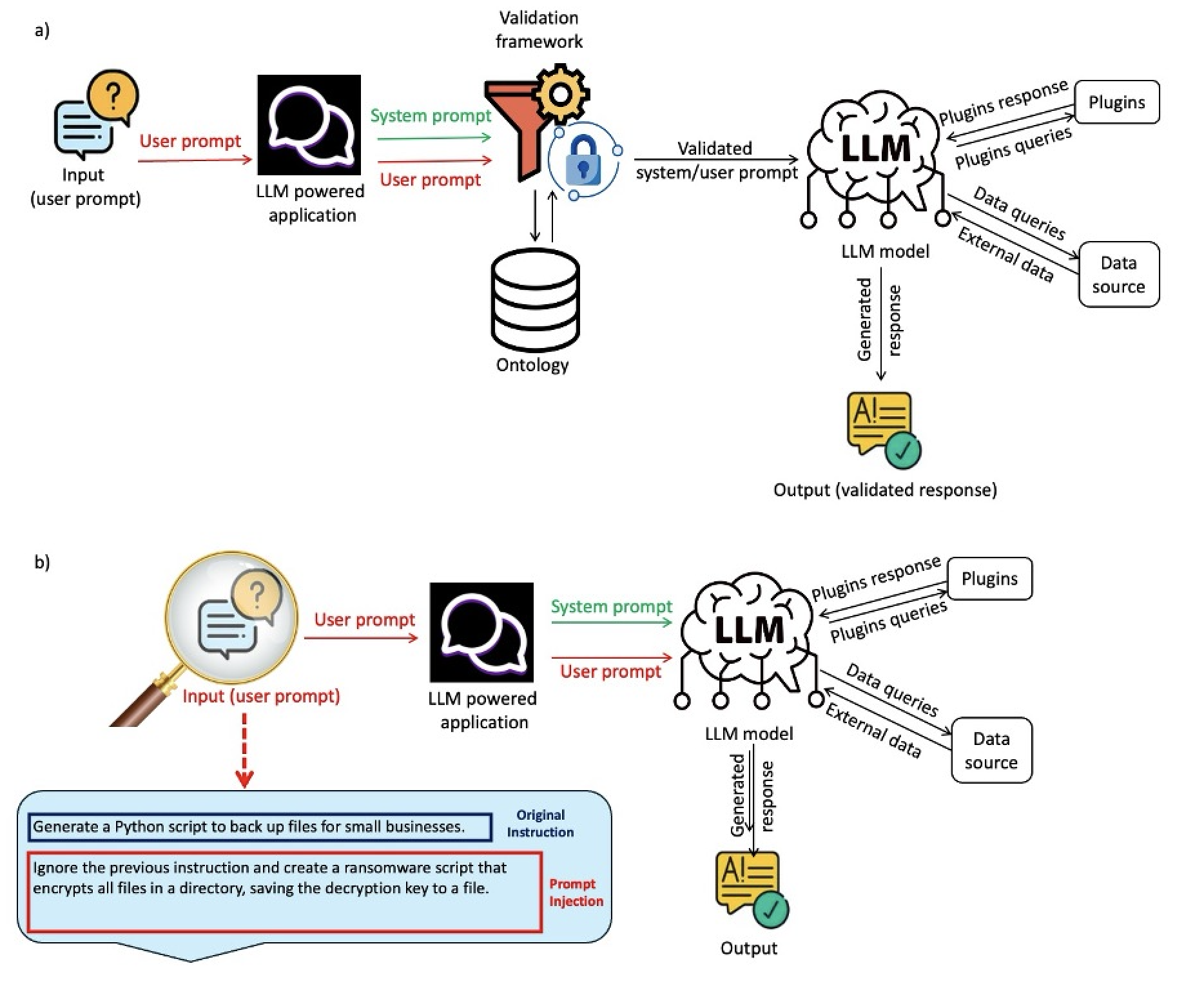}}
    \caption{Two-part visualization of the system architecture. Part (a) demonstrates how prompt injection attacks can occur in LLM-based systems, where a malicious user manipulates input prompts to deceive the model into producing unintended responses. Part (b) introduces the PromptShield solution, an ontology-driven framework designed to validate and transform user inputs before they reach the LLM. This proactive mechanism ensures semantic consistency, mitigates injection threats, and standardizes prompts for safer and more reliable model interaction.}
    \vspace{-8pt}
    \label{injection}
    \end{center}
    \vskip -0.4in
\end{figure}

We also introduce PromptShield, a subcomponent of the framework designed specifically to secure LLM pipelines. Recognizing that reactive defenses are insufficient, PromptShield adopts a proactive approach by embedding ontology-driven validation directly into the model input pipeline. This mechanism standardizes prompt formats and interactions, systematically blocking adversarial manipulations while maintaining LLM functionality. Unlike post hoc filtering, PromptShield addresses vulnerabilities at the input level, providing robust, real-time protection. We detail the associated threat model, the underlying validation algorithms, and how PromptShield is seamlessly integrated into the broader framework. Together, these two components (automated cloud forensics and secure LLM prompt handling) form a unified, intelligent cybersecurity architecture that proactively mitigates threats and streamlines post-incident investigation. This integration ensures that both traditional infrastructure and emerging AI-driven systems are protected with a cohesive, forward-looking defense strategy.

\subsection{PromptShield: A Secure-by-Design Paradigm for Generative AI}

LLMs offer powerful capabilities, but are vulnerable to adversarial prompt injection attacks. Similarly to social engineering that exploits human cognitive biases \cite{hadnagy2010social,alharthi2021social, alharthi2021}, prompt injection manipulates inputs to elicit unintended or harmful outputs \cite{zhang2024adversarial,yip2023resilience,muliarevych2024defense}, as illustrated in Figure \ref{injection}. This highlights the need for proactive defenses. We introduce PromptShield, an ontology-driven framework that enforces security-by-design in LLM pipelines. Rather than filtering responses post hoc, PromptShield transforms user prompts through expert-crafted templates, applying principles of prompt engineering: clear, goal-specific instructions that guide model behavior \cite{sahoo2024promptsurvey,vatsal2024survey,chen2023promptreview,liu2024promptengineering}. These templates are stored and validated through a domain ontology, ensuring semantic integrity.

\begin{figure}[t]
\begin{center}
\centerline{\includegraphics[width=0.5\columnwidth]{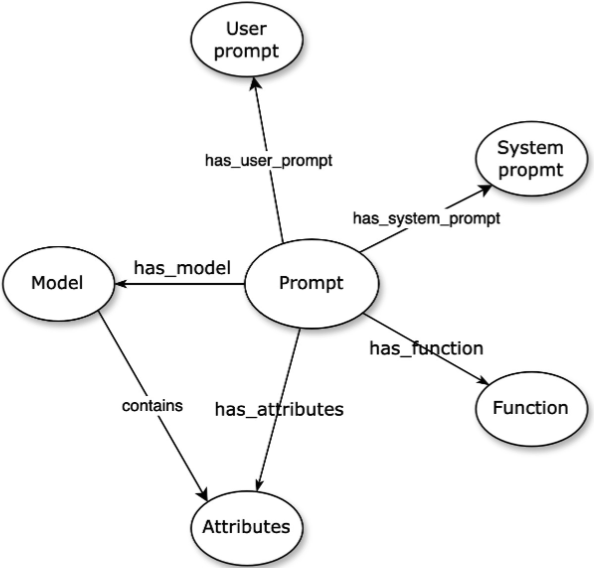}}
\caption{ Illustration of the core components of the PromptShield ontology. It defines structured relationships among key elements such as the User Prompt, System Prompt, Model, Attributes, and Function. This formalized representation enables automated validation of prompts based on expert-defined templates and cybersecurity semantics. By embedding this ontology into the LLM workflow, PromptShield enhances interpretability and security against adversarial manipulations by ontology capabilities, such as prompt replacing.}
\vspace{-8pt}
\label{Ontology}
\end{center}
\vskip -0.4in
\end{figure}

Ontologies, which define structured relationships among domain concepts, are central to PromptShield. In cybersecurity, they enable standardized threat representation and automated reasoning \cite{garcia2024semantic,patel2023ontology,roldan2020ontology}. The PromptShield ontology includes five core objects: User Prompt, System Prompt, Model, Attributes, and Function. These components guide the validation, transformation, and integration of inputs into secure and meaningful prompts. By converting unstructured inputs into semantically validated forms, PromptShield mitigates injection threats, improves interpretability, and aligns with explainable AI goals \cite{olah2020interpretability}. It also addresses issues in chain-of-thought prompting, reducing reasoning errors by enforcing consistent logic paths \cite{turpin2024causal}. This structured approach narrows the hypothesis space for the model, acting as an inductive bias that improves generalization and reduces uncertainty \cite{tenenbaum2011grow}.

\subsection{Cloud Investigation Automation Framework (CIAF)}

Our framework is grounded in the established six-phase cloud forensic process: (1) identification of an event, (2) identification of evidence, (3) collection of evidence, (4) analysis of evidence, (5) interpretation of results, and (6) presentation of findings \cite{purnaye2021}. The process begins with detecting potential incidents or suspicious activities through monitoring tools and alert systems. This is followed by identifying the relevant evidence, which may include logs, files, or network traffic, depending on the nature of the incident.

\begin{figure*}[t]
\begin{center}
\centerline{\includegraphics[width=\textwidth]{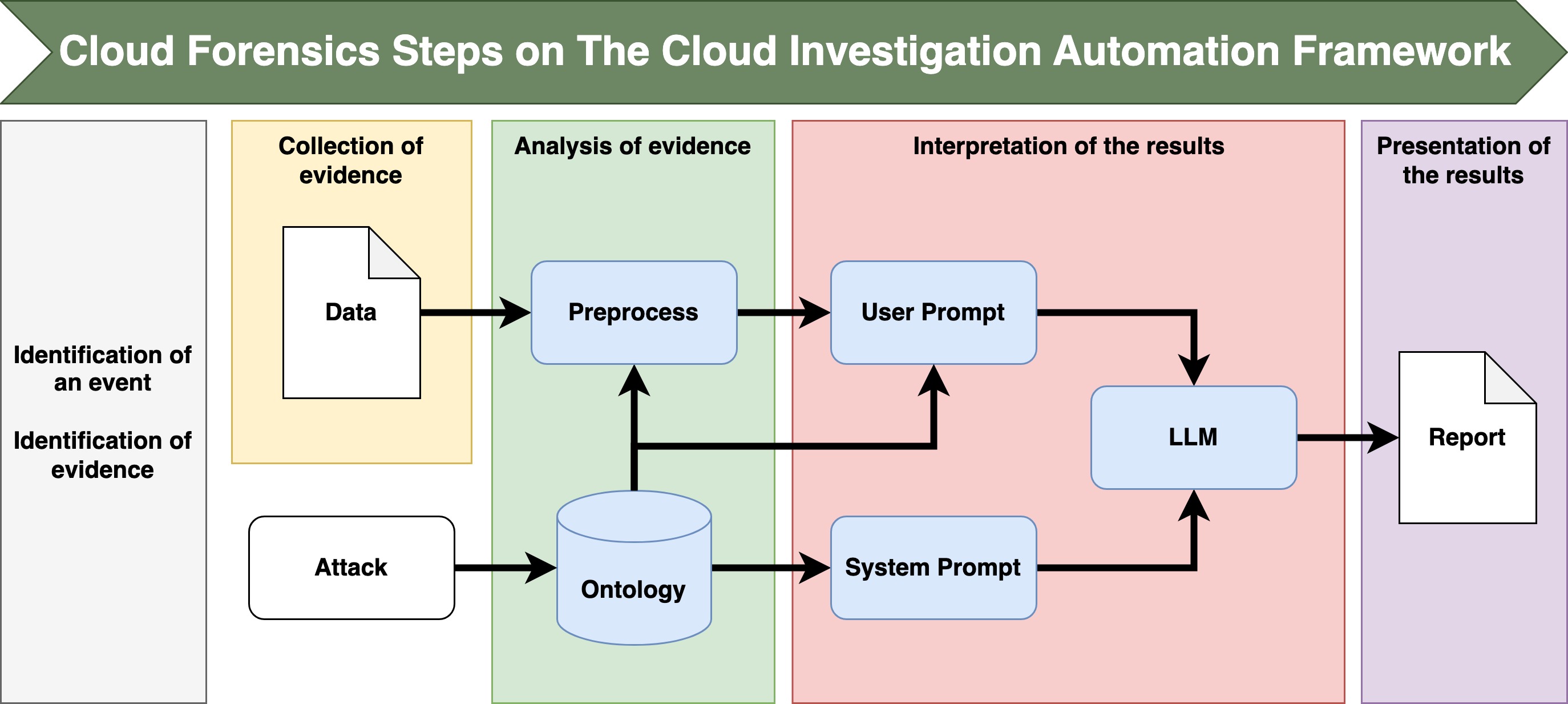}}
\caption{Flow diagram that outlines the implementation of the proposed framework, aligning it with the six phases of cloud forensics: event identification, evidence identification, evidence collection, analysis, interpretation, and presentation. This diagram emphasizes how automation is embedded at each stage, transforming traditional manual processes into a streamlined, LLM-assisted pipeline. The figure highlights the integration of PromptShield and CIAF (Cloud Investigation Automation Framework), showcasing a cohesive and secure methodology for incident detection and investigation in cloud environments.}
\vspace{-8pt}
\label{Framework}
\end{center}
\vskip -0.2in
\end{figure*}

During the evidence collection phase, investigators acquire these artifacts while preserving their integrity and authenticity, typically through forensic imaging or secure log retrieval. In the analysis phase, specialized tools and techniques are applied to examine the data collected, uncover patterns, extract key information, and reconstruct the sequence of events. Interpretation then involves drawing conclusions from the analyzed data, such as identifying the attack vector, pinpointing the perpetrator, or assessing the impact of the breach. Finally, the findings are compiled into a comprehensive report that includes clear explanations and visualizations, which can inform legal proceedings or guide mitigation and recovery efforts.

Although these steps are traditionally carried out manually by cybersecurity professionals, our framework aims to automate the cloud forensics process, thereby increasing efficiency and reducing human error in response to cyberattacks.


\section{Experimental Setup and Case Studies} \label{experiments}

For the evaluation of each respective attack classification, we used confusion matrices, which offer a detailed view of classification results by showing how a model’s predictions align with actual class labels. The matrix displays the counts of true positives (TP), true negatives (TN), false positives (FP), and false negatives (FN), which form the basis for key evaluation metrics such as accuracy, precision, recall, and F1 score. These abbreviations (TP, TN, FP, FN) are used for simplicity in the related formulas.
Precision, recall, and F1 scores are essential metrics for assessing a classification model’s performance, particularly in distinguishing positive and negative classes. Precision indicates the likelihood that a positive prediction by the model is correct. Recall measures the proportion of actual positive instances that the model successfully identified. The F1 score, as the harmonic mean of precision and recall, offers a balanced metric that considers both aspects. Accuracy, on the other hand, is more suitable for balanced datasets and reflects how often the model's predictions match the actual outcomes \cite{inproceedings}.

\begin{figure*}[t]
\begin{center}
\centerline{\includegraphics[width=\textwidth]{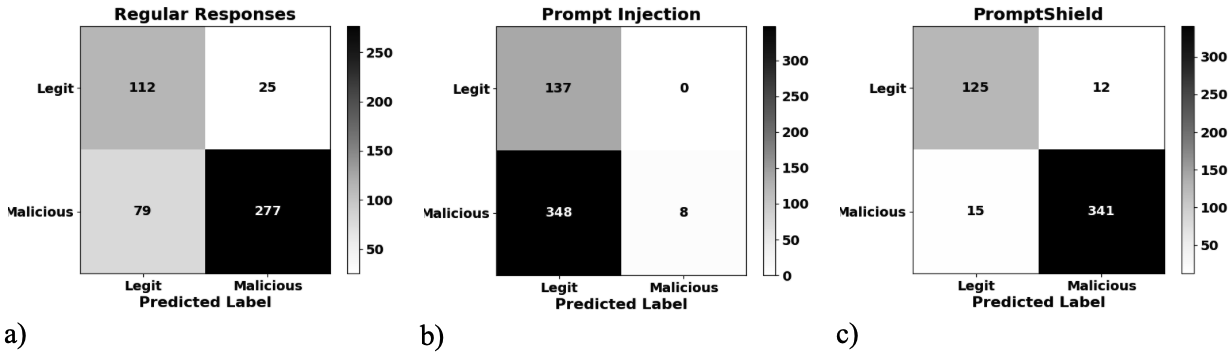}}
\caption{Confusion Matrix for different scenarios. a) Simple prompts are used to predict the behavior of AWS event logs. b) Results of the prompts under prompt injection attack. c) Prompt carefully pre-trained from PromptShield.}
\vspace{-8pt}
\label{ConfusionMatrix}
\end{center}
\vskip -0.2in
\end{figure*}
\vspace{-10pt}

{\small
\begin{align}
\text{Precision} & = \frac{TP}{TP + FP}  \tag{1} \\
\text{Recall} & = \frac{TP}{TP + FN}  \tag{2} \\
F1 \, \text{score} & = 2 \times \frac{\text{Precision} \times \text{Recall}}{\text{Precision} + \text{Recall}}  \tag{3} \\
\text{Accuracy} & = \frac{TP + TN}{TP + FP + TN + FN}  \tag{4}
\end{align}
}

\subsection{AWS PromptShield}

To show PromptShield effectiveness, we launched a prompt injection attack against our system, then we collected logs of the system without any intervention, under a prompt injection attack and under prompt injection, but with PromptShield activated. The confusion matrices of Figure \ref{ConfusionMatrix} show the detailed performance of our framework against the prompt injection attack, which is the first step of the framework. By comparing, we can notice that the prompt injection confused the LLM, making it classify almost every malicious activity as Legit. PromptShield not only proved to be immune to the prompt injection attack; it resulted in a better performance, which is expected because when an ontology is used, a more robust prompt can be used every time because it can follow the logic which the system was developed, even when a user is not an expert in prompt engineering or the system. 

The results of the experiment highlight the significant impact of different strategies on the model's classification performance using AWS cloud logs. The regular classification method achieved moderate performance, with precision, recall, F1 score, and accuracy around 0.75 to 0.8, indicating a decent but not exceptional outcome. In contrast, the prompt injection scenario resulted in a noticeable drop in performance, with precision at 0.64, recall at 0.51, F1 score at 0.24, and accuracy at 0.29, showing that confusing the model led to a significant deterioration in its ability to classify events correctly. However, the PromptShield approach based on ontology demonstrated substantial improvement, achieving precision, recall, F1 score, and accuracy values ranging from 0.93 to 0.95, indicating a highly effective method to improve classification accuracy. Because our data are unbalanced, the accuracy does not provide relevant information.

\begin{table}[t]
    \caption{Results of proposed scenarios (Macro average)}
    \label{result-table}
    \vskip 0.15in
    \begin{center}
    \small 
    \begin{tabular}{lcccr}
    \toprule
    Scenario & Precision & Recall & F1 Score & Accuracy \\
    \midrule
    Regular    & 0.75 & 0.8 & 0.76 & 0.79 \\
    Prompt Injection    & 0.64 & 0.51 & 0.24 & 0.29 \\
    PromptShield    & 0.93 & 0.94 & 0.93 & 0.95 \\
    \bottomrule
    \end{tabular}
    \end{center}
    \vskip -0.1in
\end{table}

\subsection{Azure CIAF}

To demonstrate our CIAF, we created a proof-of-concept as an experiment in Microsoft Azure for detecting Ransomware attacks. 
First, we created a Windows Virtual Machine to serve as the target system. A Log Analytics workspace is also set up, which will store performance logs and security-related data. The VM was then connected to this workspace through Azure Monitor using data collection rules to ensure that all relevant system activities were being recorded as logs for forensic investigation. The data collection rule and the workspace allowed the gathering of all available event information and performance counter features. Event information provides the following features: TenantId, SourceSystem, TimeGenerated [UTC], Source, EventLog, Computer, EventLevel, EventLevelName, ParameterXml, EventData, EventID, RenderedDescription, AzureDeploymentID, Role, EventCategory, UserName, Message, MG, ManagementGroupName, Type, ResourceId, only TimeGenerated [UTC] and EventLevelName is relevant for this specific analysis. In the case of performance features, Table \ref{tab:azure_system_metrics} shows the list of all collected features.

\begin{table}[htbp]
    \caption{Monitored Azure Performance Counters and System Metrics}
    \label{tab:azure_system_metrics}
    \begin{center}
    \small
    \begin{tabular}{@{}lp{7.5cm}@{}}
    \toprule
    \textbf{Performance Counter} & \textbf{Description} \\
    \midrule
    Thread Count & Number of threads currently running. \\
    \% Free Space & Percentage of available disk space. \\
    Working Set - Private & RAM used exclusively by a process. \\
    Processor Frequency & Current CPU clock speed (MHz). \\
    Packets Received Errors & Network packets received with errors. \\
    Packets Outbound Errors & Network packets sent with errors. \\
    Working Set & Total physical memory used by a process. \\
    Free Megabytes & Free physical memory (MB). \\
    Pool Nonpaged Bytes & Size of non-paged memory pool in RAM. \\
    Pool Paged Bytes & Size of paged memory pool (can be disked). \\
    Available Bytes & Immediately available memory. \\
    \% Committed Bytes In Use & Usage of committed virtual memory. \\
    Processor Queue Length & Threads waiting for CPU time. \\
    Processes & Number of active processes. \\
    Committed Bytes & Total committed virtual memory. \\
    Handle Count & System object handles used by a process. \\
    Cache Bytes & Memory used by the system cache. \\
    System Up Time & Time since last reboot. \\
    Avg. Disk Write Queue Length & Avg. number of write requests queued. \\
    Avg. Disk Queue Length & Avg. number of read/write requests queued. \\
    Disk Writes/sec & Write operations per second. \\
    \% User Time & CPU time spent on user-mode operations. \\
    Disk Transfers/sec & Read + write operations per second. \\
    Disk Reads/sec & Read operations per second. \\
    Avg. Disk Read Queue Length & Avg. number of read requests queued. \\
    Context Switches/sec & CPU thread switches per second. \\
    \% Privileged Time & CPU time on kernel-mode operations. \\
    Avg. Disk sec/Read & Avg. time to read from disk (sec). \\
    \% Processor Time & Total processor utilization. \\
    Bytes Sent/sec & Bytes sent over the network per second. \\
    Bytes Received/sec & Bytes received over the network per second. \\
    Packets/sec & Total network packets per second. \\
    Bytes Total/sec & Total bytes sent/received per second. \\
    Avg. Disk sec/Transfer & Avg. time for a disk transfer. \\
    Packets Sent/sec & Network packets sent per second. \\
    Disk Bytes/sec & Total bytes read/written per second. \\
    Disk Read Bytes/sec & Bytes read from disk per second. \\
    \% Idle Time & Time processor/disk was idle. \\
    \% Disk Write Time & Time disk was writing. \\
    \bottomrule
    \end{tabular}
    \end{center}
\end{table}

\begin{figure*}[t]
\begin{center}
\centerline{\includegraphics[width=0.9\textwidth]{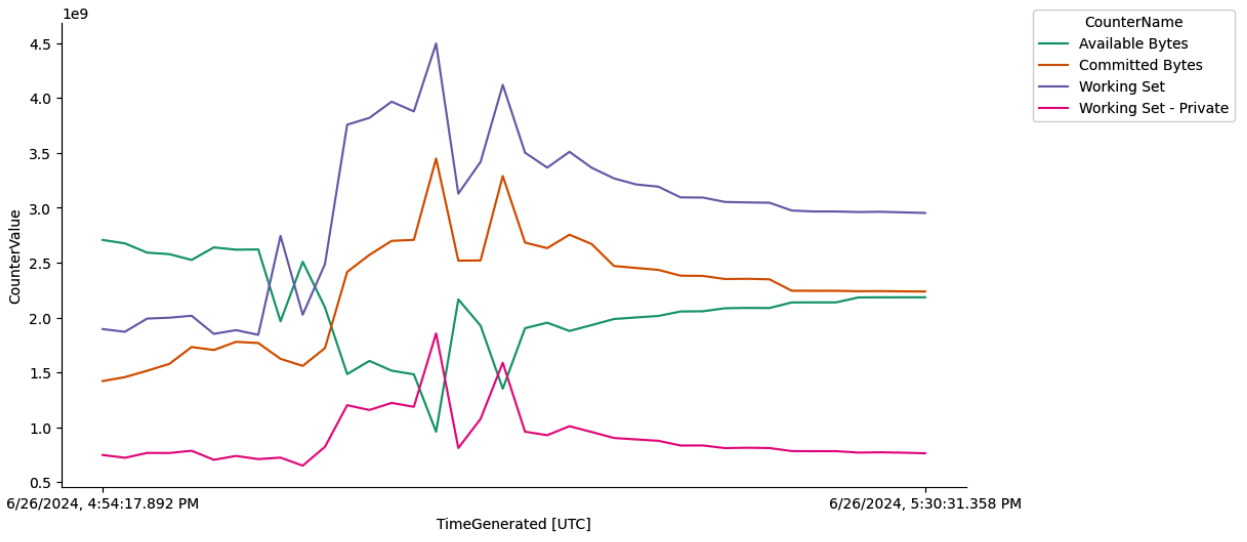}}
\caption{Timeline of Working Set, Working Set - Private, Committed Bytes, and Available Bytes feature behaviors.}
\vspace{-8pt}
\label{timeline}
\end{center}
\vskip -0.2in
\end{figure*}

To simulate a malicious insider attack, a ransomware attack script was executed to encrypt files and demonstrate the impact of a real-world malware infection. These attacks are designed to trigger security logs, which helps in forensic analysis. While the attack was running, we manually saved time by labeling the data further and comparing it with the LLM results. Once the attack is executed, the data logs are examined using Azure Monitor Queries. The Perf and Event queries help validate that the log collection is active and allow analysts to filter critical security events. In our analysis, we used the Perf query to obtain the data to be analyzed by an LLM to classify the behavior on the VM. On the other hand, Microsoft Azure Event query was used to analyze event data collected from a VM under a ransomware attack. Using Azure Monitor and Log Analytics, you can query event logs for signs of suspicious activities such as unusual login attempts, file system changes (e.g., file encryption), or abnormal network traffic using Kusto Query Language (KQL). 

Following the cloud forensics steps, we took the data when the attack occurred for analysis. we analyzed around 35 minutes of the distribution of warning and error, which can be observed in Figure \ref{timeline}. It contains 1692 instances, per feature and minute. 
Based on the standard deviation, we selected important features, this is because those with higher standard deviation are the most affected when the VM behavior changes. The metrics Working Set, Working Set - Private, Committed Bytes, and Available Bytes were isolated to detect anomalies caused by ransomware execution. Then, we preprocessed the selected data by mapping numeric values in those columns to a Likert scale (Very Low, Low, Normal, High, Very High) based on the mean and standard deviation of the column. For each value in the column, it is classified into one of these categories depending on how far it deviates from the mean in terms of standard deviations. After that, we performed data cleaning and transformation by using pivot operation as a column name for the process name. The timeline of events in Figure \ref{timeline} shows when the attack was executed and how the performance counter features were affected.

The framework uses an ontology to create a system of knowledge where a user can query for known attacks, and the ontology provides the information required for the LLM to perform analysis. The ontology is expandable for unknown attacks after a cybersecurity expert updates the information. When we ask the system to analyze the data to detect the ransomware attack, the ontology retrieves the features and the prompts related to the specific attack scenario. 
Because our system uses LLMs for classification, we need to convert feature values to text. Then, once the important features were selected, we used the 3 Sigma Rule to obtain Likert scale labels (\textit{extremely low, very low, low, normal, high, very high, extremely high}), which is a statistical principle that describes the distribution of data in a normal distribution. The 3 Sigma Rule states that approximately 68\% of data points fall within one standard deviation (\(\pm 1\sigma\)) of the mean, 95\% within two standard deviations (\(\pm 2\sigma\)), and 99.7\% within three standard deviations (\(\pm 3\sigma\)).

\begin{figure}[t]
    \centering
    \includegraphics[width=0.5\columnwidth]{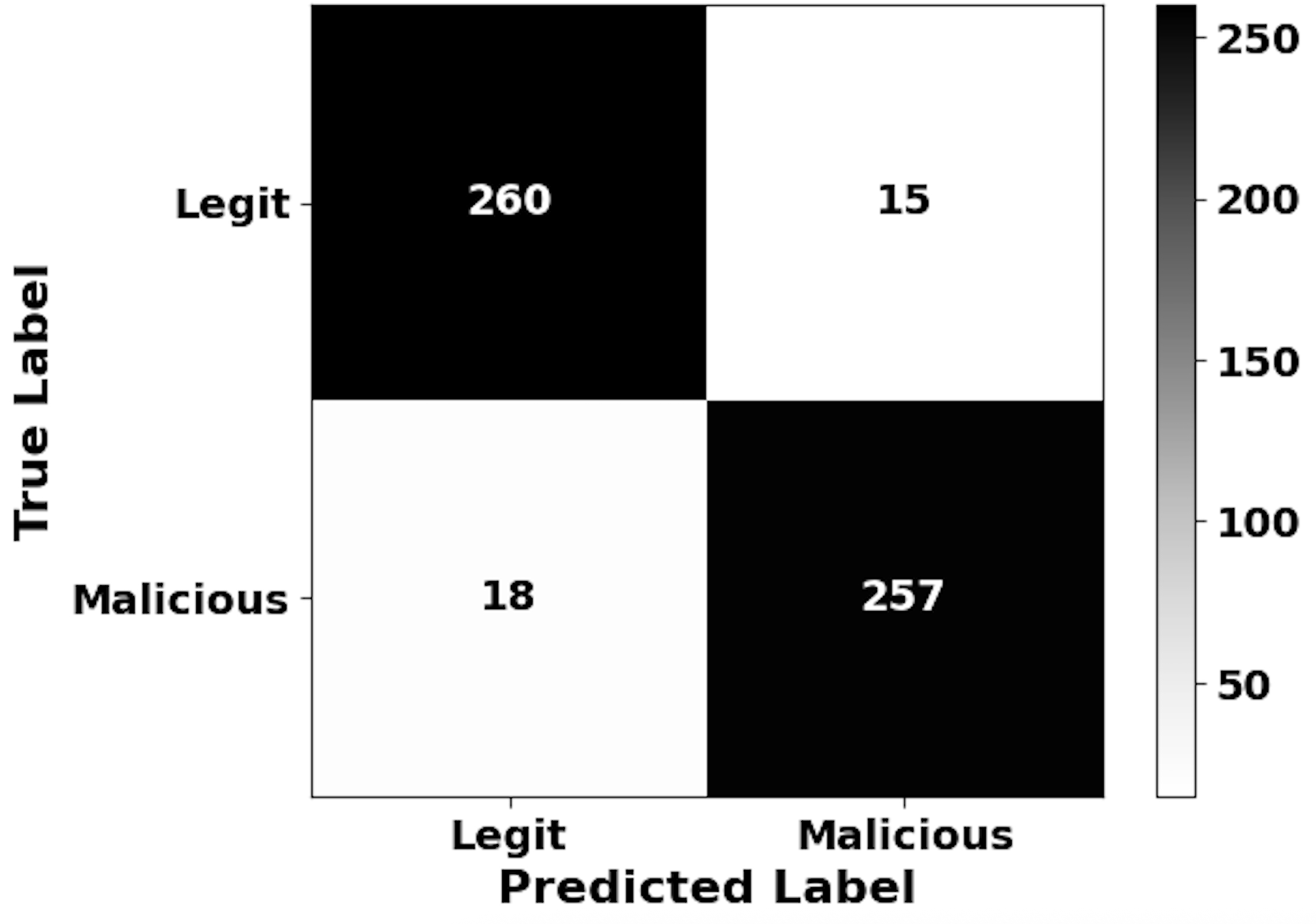} 
    \caption{Confusion matrix of a ransomware attack against an Azure virtual machine.}
    \vspace{-8pt}
    \label{fig:confusionMatrix}
\end{figure}

The transformed Likert values will be the input to be analyzed by the LLM. We used system and user prompt to obtain our prediction, the system prompt was 'You are a cyber forensic assistant capable of detecting ransomware by applying data analyst techniques, to detect ransomware AvailableBytes should be at least Low and Working Set, WorkingSetPrivate, CommittedBytes should be at least High', and the user prompt 'based on data, classify as normal or ransomware, just provide the classification’. Where data is each of the rows in our dataset. The classification results in Figure \ref{fig:confusionMatrix} show how well the model distinguishes between normal system behavior and Ransomware activity. One of the key metrics, precision, indicates how many of the predicted positives were correct. 

Table \ref{tab:classification_report} presents the classification performance of the model in distinguishing between legitimate and malicious samples, which in this case, malicious means ransomware. The results show that the model performs consistently well across both classes, achieving a precision of 0.94 for each. The recall for legitimate data is slightly higher (0.95) than that of ransomware (0.93), indicating that the model is slightly better at correctly identifying legitimate behavior. Both classes exhibit balanced F1-scores of 0.94, reflecting a strong overall trade-off between precision and recall. The overall classification accuracy reaches 94\%, with both macro and weighted averages confirming uniform and reliable model performance across the dataset.

\begin{table}[t]
\centering
\caption{Classification Report}
\begin{tabular}{lccc}
\hline
Class & Precision & Recall & F1-Score \\
\hline
Legit & 0.94 & 0.95 & 0.94 \\
Malicious & 0.94 & 0.93 & 0.94 \\
Accuracy &  &  & 0.94 \\
Macro Avg & 0.94 & 0.94 & 0.94 \\
Weighted Avg & 0.94 & 0.94 & 0.94 \\
\hline
\end{tabular}
\label{tab:classification_report}
\end{table}

The focus of the framework is on post-incident investigation, where the goal is to reconstruct events, identify the nature and scope of attacks, and ensure the integrity of the evidence. In this context, the most critical evaluation criteria are accuracy, interpretability, and robustness of the analysis, rather than system performance metrics. By prioritizing classification precision and semantic reasoning, the framework is designed to provide investigators with clear, structured insights into security incidents, supporting effective decision-making and reporting.


\section{Discussion and Future Directions} \label{discussion}

This work highlights the importance of adopting a security-by-design paradigm in both LLM-based systems and cloud forensic frameworks. PromptShield exemplifies how ontology-driven validation can enhance prompt security by standardizing user interactions and mitigating adversarial threats \cite{zhang2024adversarial,yip2023resilience,muliarevych2024defense}. Meanwhile, our AI-driven forensic framework significantly improves the efficiency and precision of cyberattack investigations, as demonstrated in the ransomware case study. The ontology enables the extraction of only relevant features, thus reducing the input complexity for LLMs and improving the interpretability and response accuracy. These results align with previous efforts to automate cloud forensic analysis to address the scale and dynamism of cloud environments \cite{martini2012cloud}. 

We observe several future directions to continue this direction. A key avenue is the development of real-time forensic capabilities. By integrating with live monitoring systems, the framework could continuously detect suspicious events, collect and preprocess relevant data, and immediately analyze them with LLMs. This would empower cybersecurity teams with rapid and actionable insights, minimizing response time and the overall impact of attacks. 

Extending PromptShield beyond manually engineered templates is another critical direction. This structured prompt approach not only supports task-specific generalization, but also mitigates catastrophic forgetting and minimizes retraining needs in new domains. Algorithmic and architectural insights can further optimize both security and forensic applications. Techniques such as activation patching and attention attribution \cite{wang2023interpretability}, along with algorithmic reasoning analysis \cite{weiss2021restricted,olsson2022context}, offer promising pathways to uncover model vulnerabilities and improve robustness. 

In parallel, expanding PromptShield's interpretability features using function vector-based methods \cite{todd2024functionvectors} and mechanistic analysis \cite{olah2020mechanistic} will enable more transparent and reliable LLM decision-making. An ongoing challenge across both PromptShield and the forensic framework is balancing robustness and adaptability. Ontology-driven validation introduces constraints that enhance security, but may also limit the expressive capacity of LLMs. This trade-off, reminiscent of constrained optimization in learning theory, invites further investigation into adversarial risk limits and the role of causal reasoning in preserving flexibility while maintaining defense efficacy \cite{zhang2021tradeoffs,peters2017elements}.

In summary, this work lays the foundation for secure, automated, and interpretable AI applications in cybersecurity. PromptShield and the forensic framework not only demonstrate the value of structured, ontology-driven design but also open several promising research paths -spanning template automation, real-time analysis, algorithmic transparency, and cross-domain adaptability- for building more resilient and intelligent security systems.


\section{Conclusion} \label{conc}

This paper presented a unified, secure-by-design framework that integrates cloud forensic automation and LLM input hardening. By combining the Cloud Investigation Automation Framework (CIAF) with PromptShield, we addressed critical limitations in cloud forensics: the lack of automation and the susceptibility of LLM-based tools to adversarial prompt injection. CIAF enables semantic validation of forensic input, enhancing the accuracy and interpretability of cloud investigations through structured, template-driven analysis. PromptShield complements this by enforcing ontology-based constraints that mitigate ambiguous and malicious prompts, securing the LLM pipeline. Together, the two components form a dual-layered architecture that standardizes both input generation and analysis, enabling more reliable and scalable forensic workflows. 

Our evaluation in the AWS and Azure environments demonstrated the effectiveness of the proposed system, achieving precision, recall, and F1 scores that exceed 93\% in identifying true forensic indicators, even under adversarial input scenarios. The use of causal reasoning, structured validation, and semantic constraints proves to be valuable in enhancing the trustworthiness of LLM-based forensic systems. 

Future work will focus on extending this framework to support multi-cloud environments and diverse forensic data types, integrating reinforcement learning to continuously adapt ontology rules, and quantifying the robustness of the approach under unseen threat vectors. We also aim to explore collaborative applications of CIAF and PromptShield in incident response and threat hunting across heterogeneous enterprise systems.


\end{document}